\documentclass[11pt,a4paper]{article}
\usepackage{amsmath,amssymb,titling,authblk}
\usepackage[top=2.3cm,right=2.3cm,left=2.3cm,bottom=2.3cm]{geometry}
\usepackage{graphicx}
\usepackage{color} 
\usepackage{slashed}
\usepackage[pdfstartview=FitH,colorlinks=true,linkcolor=blue,anchorcolor=red,citecolor=magenta,urlcolor=blue]{hyperref}
\usepackage{amscd}
\usepackage[normalem]{ulem}
\usepackage{appendix}
\usepackage{bbold}
\usepackage{mathrsfs}
\usepackage{pdfsync}
\usepackage{bbm}
\usepackage{bm}
\usepackage[arrow,matrix,curve]{xy}
\usepackage{bbding}
\usepackage{wasysym}
\usepackage{booktabs}
\usepackage{siunitx}

\usepackage{epsf}
\usepackage{epsfig}
\usepackage{wrapfig}
\usepackage[utf8]{inputenc}
\usepackage{subcaption}
\usepackage{cite}
\DeclareCaptionFormat{custom}
{%
    \textbf{#1#2}\textit{\small #3}
}
\allowdisplaybreaks[4]
\numberwithin{equation}{section}

\definecolor{verde}{cmyk}{.83,.21,1,.08}
\definecolor{darkorchid}{rgb}{0.6, 0.2, 0.8}
\definecolor{darkgreen}{rgb}{0,.5,0}

\def\({\left(}
\def\){\right)}
\def\[{\left[}
\def\]{\right]}

\newcommand{\ii}{\mathrm{i}}

\newcommand{\dd}{\mathrm{d}}

\newcommand{\be}{\begin{equation}}
\newcommand{\ee}{\end{equation}}
\newcommand{\bea}{\begin{eqnarray}}
\newcommand{\eea}{\end{eqnarray}}
\newcommand{\del}{\partial}

\newcommand{\la}{\label}

\begin{document}
\author[1,2]{M. A. Kurkov}
\affil[1]{\textit{\footnotesize Dipartimento di Fisica ``E. Pancini'', Universit\`a di Napoli Federico II, Complesso Universitario di Monte S. Angelo Edificio 6, via Cintia, 80126 Napoli, Italy.}}
\affil[2]{\textit{\footnotesize INFN-Sezione di Napoli, Complesso Universitario di Monte S. Angelo Edificio 6, via Cintia, 80126 Napoli, Italy.}}
\affil[ ]{\footnotesize e-mail: \texttt{ max.kurkov@gmail.com}}
\title{	Light Propagation in $\kappa$-Minkowski Space-Time: Gauge Ambiguities and Invariance}
\maketitle
\begin{abstract}\noindent
We study the noncommutative $U(1)$ gauge theory on the $\kappa$-Minkowski space-time at the semiclassical approximation. We construct exact solutions of the deformed Maxwell equations in vacuum, describing localized signals propagating in a given direction. The propagation velocity appears to be arbitrary. We figure out that the wave packets with different values of the propagation velocity are related by noncommutative gauge transformations. Moreover, we show that spatial distances between particles are gauge-dependent as well. We explain how these two gauge dependencies compensate each other, recovering gauge invariance of measurement results. According to our analysis, the gauge ambiguity of the speed of light can be absorbed into a redefinition of the unit of length and, therefore, cannot be measured experimentally. 
\end{abstract}

\section{Introduction}
String theory~\cite{Seiberg:1999vs} and other approaches to quantum gravity~\cite{Freidel:2005me} suggest that at small length scales space-time exhibits a noncommutative geometric structure~\cite{Doplicher:1994tu}. The dynamics of gauge fields and particles in noncommutative space-time is described by noncommutative gauge theory~\cite{Szabo}. For a comprehensive review, we refer to~\cite{Hersent:2022gry}. The present article is devoted to noncommutative electrodynamics, that is, to the noncommutative $U(1)$ gauge theory.

Consider an infinitesimal gauge transformation $\delta_f$  of some dynamical variable, e.g., of the  {gauge} potential $A$,
\be
A \longrightarrow \tilde{A} := A + \delta_f A,
\ee
where $f$ denotes the gauge parameter. While in the usual $U(1)$ gauge theory, these transformations commute\footnote{ {Here and in Eq.~\eqref{NCalg} we use the superscripts $\mathbf{0}$ and $\mathrm{NC}$ to distinguish the commutative $U(1)$ gauge variations from the (full) noncommutative ones.}}:
\be
\[\delta^{ {\mathbf{0}}}_f,\delta_g^{ {\mathbf{0}}}\] = 0,  \la{Calg}
\ee
the space-time noncommutativity gives rise to the \emph{deformed} gauge algebra:
\be
\[\delta_f^{ {\mathrm{NC}}},\delta_g^{ {\mathrm{NC}}}\] = \delta^{ {\mathrm{NC}}}_{-\ii [f,g]_{\star}}, \la{NCalg}
\ee
which is essentially non-Abelian. The expression
\be
 [f,g]_{\star} := f\star g - g\star f,
\ee 
is called the star-commutator of the gauge parameters $f$ and $g$, and the symbol $\star$ denotes the Kontsevich star-product, characterizing the space-time noncommutativity.
In the novel approach to noncommutative $U(1)$ gauge theory proposed in~ {\cite{Kupriyanov:2019ezf,Kupriyanov:2019cug,Kupriyanov:2020sgx}}, the algebra~\eqref{NCalg} is taken as a fundamental postulate. 
 
Being a non-local field theory, the noncommutative electrodynamics is extremely complicated from a technical point of view. However, in the semiclassical approximation, things become much simpler.
In this approximation, the star-commutator $[\,,\,]_{\star}$ reduces to the Poisson bracket $\{\, ,\,\}$ times the imaginary unit $\ii$. Therefore, the non-Abelian nature of the algebra~\eqref{NCalg} is preserved: 
\be
\[\delta_f,\delta_g\] = \delta_{\{f,g\}},   \la{LPalgZero}
\ee
although the field-theoretical model becomes \emph{local}. For Lie-algebra-type noncommutativities, this local semiclassical framework is called  \emph{Lie-Poisson electrodynamics}, or Lie-Poisson gauge theory  {(hereafter, $\delta_f$ denotes infinitesimal Lie-Poisson gauge transformations). The first steps towards the Lie–Poisson gauge formalism were made in~\cite{Kupriyanov:2019ezf}, where the algebra \eqref{LPalgZero} was considered for the $\mathfrak{su}(2)$ case.} 

In recent years, the Lie-Poisson approach has attracted significant attention~\cite{Kupriyanov:2019ezf, Kupriyanov:2019cug, Kupriyanov:2020sgx,Kupriyanov:2020axe,Kupriyanov:2021aet,Kupriyanov:2021cws,Abla:2022wfz,Kupriyanov:2022ohu,Abla:2023odq,Kupriyanov:2023gjj,Kupriyanov:2023qot,DiCosmo:2023wth,Bascone:2024mxs,Kupriyanov:2024dny,Sharapov:2024bbu,Basilio:2024bir,Abla:2024wtr}, and the present work continues the research line of the quoted references\footnote{Of course, there are other models studied in the context of the semiclassical limit of noncommutative gauge theories. A detailed comparison of our approach with others is presented in~\cite{Kupriyanov:2023gjj}.}.

The semiclassical approximation allows for the idealization of a point-like particle, which would be strictly forbidden in the full noncommutative setting due to the uncertainty relations between coordinates. The local interaction between point-like particles and the gauge field within the Lie-Poisson electrodynamics was studied in detail in~\cite{Basilio:2024bir}. As we shall see below, the concept of a point-like particle is very useful for the understanding of various subtle aspects of the formalism.
  
In the present paper, we focus on the Lie-Poisson electrodynamics with the $\kappa$-Minkowski noncommutativity. This noncommutativity and its variations have been broadly studied in the literature, see, e.g.,~\cite{Lizzi:2020tci,Mathieu:2020ywc,Lizzi:2021rlb,Mathieu:2021mxl,Meljanac:2022qed,Carotenuto:2020vnv,Fabiano:2023xke} for recent progress\footnote{For a detailed list of references, we refer to the review~\cite{Hersent:2022gry}.}.  Five-dimensional noncommutative gauge theories in this context were considered in~\cite{Mathieu:2020ywc,Mathieu:2021mxl}. In the scope of the Lie-Poisson electrodynamics, the deformed noncommutative four-dimensional Maxwell equations were obtained in~\cite{Kupriyanov:2023gjj}. 
According to~\cite{Kupriyanov:2023gjj}, these deformed field equations exhibit plane wave solutions
with the unusual dispersion relation:
\be
\big(1-\kappa^2\,\varepsilon^2\big) \,k_0^2 -\big|\vec{k}\big|^2 = 0, \la{drL}
\ee 
which involves the deformation parameter\footnote{In~\cite{Kupriyanov:2023gjj} the deformation parameter is denoted by $\lambda$.}  $\kappa$ and the amplitude $\varepsilon$ of the plane wave.
We raise the natural question of what the speed of light in this model is. While plane waves provide information about the phase velocity only, a study of the speed of light requires an analysis of localized signals or wave packets.

The rest of this paper is organized as follows. In Sec.~\ref{formalism}, we describe the main notions of the Lie-Poisson gauge formalism. 
The dynamical $\kappa$-Minkowski model proposed in~\cite{Kupriyanov:2023gjj} is briefly reviewed in Sec.~\ref{kappasetup}. The subsequent three sections contain our new results. In particular, in Sec.~\ref{SOLU} we present exact solutions describing travelling wave packets in the deformed setting. In Sec.~\ref{GT}, we establish the gauge dependence of their propagation velocity. In Sec.~\ref{WGO}, we show that the spatial distances between particles are gauge-dependent as well. We explain how the two gauge dependencies compensate each other, thereby restoring the gauge invariance of measurement results. We round up with a short summary of our findings and concluding remarks in Sec.~\ref{conclu}. 

\section{Basic elements of the Lie-Poisson gauge formalism}
\la{formalism}
Consider a $d$-dimensional manifold $\mathcal{M}\simeq \mathbb{R}^d$ representing space-time. The local coordinates on $\mathcal{M}$ are denoted by $x^\mu$. Greek indices $\mu$, $\nu$, $\dots$ run from $0$ to $d-1$ . We assume that $\mathcal{M}$ is equipped with a Poisson bivector $\Theta$  of Lie algebraic type,
\be
\Theta^{\mu\nu}(x) = f^{\mu\nu}_{\xi} x^{\xi},  \la{ThetaDef}
\ee
where  $f^{\mu\nu}_\xi$ are the structure constants of a given $d$-dimensional Lie algebra $\mathfrak{g}$. The corresponding Lie group, which is unique up to a covering, will be addressed as $G$.

Now, consider the Poisson bracket associated with the Poisson bivector~\eqref{ThetaDef},
\be
\{f,g\} = \Theta^{\mu\nu} \,\partial_\mu f\, \partial_\nu g ,\qquad \forall f,g \in \mathcal{C}^{\infty}(\mathcal{M}). \la{pbr}
\ee
By definition, Lie-Poisson electrodynamics is a non-Abelian deformation of the $U(1)$ gauge theory,
where the infinitesimal gauge transformations close the Lie-Poisson gauge algebra:
\be
\[\delta_f,\delta_g\] = \delta_{\{f,g\}},   \la{LPalg}
\ee
mentioned in the Introduction. As we wrote above, this field-theoretical model describes the semiclassical approximation of  noncommutative electrodynamics with the Lie-algebra-type noncommutativity,
\be
\[x^{\mu},x^{\nu}\]_{\star} =  \ii \,f^{\mu\nu}_{\xi} x^{\xi}.
\ee

At the limit $\Theta\to0$, the Poisson bracket~\eqref{pbr} vanishes, and the non-Abelian algebra~\eqref{LPalg} reduces to its Abelian counterpart. Thus, in what follows this limit is referred to as the \emph{commutative limit}.

The main constituents of the Lie-Poisson gauge formalism can be defined in terms of differential
geometric structures on the Lie group $G$ . Let $\gamma^\mu$ and $\rho_\nu$  be bases of left-invariant vector fields and
right-invariant one-forms on $G$, respectively. We also introduce local coordinates $p_\mu$  on  $G$, such that
\be
\lim_{\Theta\to0} \gamma^\mu_\nu(p) = \delta^\mu_\nu,\qquad \lim_{\Theta\to0} \rho^\mu_\nu(p) = \delta^\mu_\nu. \la{GRcomLim}
\ee
In these expressions, $\gamma^\mu_\nu (p)$ and $\rho^\mu_\nu(p)$ denote the components of $\gamma^\mu$ and $\rho_\nu$ in the natural bases
$\frac{\partial}{\partial p_\nu}$ and $\dd p_\mu$, respectively,
\be
\gamma^\mu(p) =  \gamma^\mu_\nu (p)\,\frac{\partial}{\partial p_\nu}, \qquad \qquad \rho_\nu (p) = \rho^\mu_\nu(p)\, \dd p_\mu.
\ee
 {Throughout this paper, we assume that the group $G$ is covered by a single coordinate chart, and we denote by $g(p)$ the element of $G$ whose local coordinates in this chart are $p\in\mathbb{R}^d$. As we shall see below, for the $\kappa$-Minkowski case this assumption holds.
}

The infinitesimal gauge transformations of the  {gauge field components} $A_{\mu}(x)$, which close the Lie-Poisson gauge algebra~\eqref{LPalg}, read~\cite{ Kupriyanov:2019cug,Kupriyanov:2020sgx}:
\be
\delta_f A_\mu(x) =\gamma^r_\mu(A(x))\,\partial_r f(x) +\{A_\mu(x) ,f(x)\}.  \la{igt}
\ee
The deformed field strength, which transforms in a gauge-covariant manner under these transformations,
\be
\delta_f {\cal F}_{\mu\nu}(x)=\{{\cal F}_{\mu\nu}(x),f\}, \la{igtF}
\ee
is defined as follows~\cite{Kupriyanov:2021aet}:
\be
\mathcal{F}_{\mu\nu}(x) := \rho_\mu^\xi(A)\,\rho_\nu^\sigma(A)\big(\gamma_\xi^\zeta(A)\,\partial_\zeta A_\sigma-\gamma_\sigma^\zeta (A)\,\partial_\zeta A_\xi+\{A_\xi  {,}A_\sigma\}\big). \la{Fstr}
\ee 

Finally, the deformed gauge-covariant derivative is given by~\cite{Kupriyanov:2021aet}
\be
{\cal D}_\mu\psi(x) :=\rho_\mu^\nu(A)\, (\gamma_\nu^\xi(A) \del_\xi \psi+\{A_\nu,\psi\}), \qquad \forall \psi \in \mathcal{C}^{\infty}(\mathcal{M}). \la{gcodder}
\ee
This operator is constructed in such a way that for any field  $\psi$ transforming covariantly under the
gauge transformations~\eqref{igt},
 \be
 \delta_f \psi =\{ \psi,f\} ,
 \ee
the expression $ {\cal D}_\mu\psi $ transforms in a gauge-covariant manner as well:
\be
\delta_f\left({\cal D}_\mu\psi\right)=\{{\cal D}_\mu\psi,f\}.
\ee
 {Note that in the references~\cite{Kupriyanov:2019cug,Kupriyanov:2020sgx,Kupriyanov:2021aet}, the quantities $\gamma$ and $\rho$ were introduced as solutions of certain partial differential equations. Their differential-geometric interpretation in terms of left-invariant vector fields and right-invariant one-forms on $G$, adopted in the present paper, was clarified later in~\cite{Kupriyanov:2023qot,Sharapov:2024bbu,Kupriyanov:2024dny}.}

\section{ $\kappa$-Minkowski setup}\label{firstres}
\la{kappasetup}
Below, we focus on the  $\kappa$-Minkowski noncommutativity at $d=4$. The non-zero Poisson brackets between the space-time coordinates read:
\be
\{x^0, x^j\} = \kappa \, x^j   = - \{x^j, x^0\}
,\qquad j=1,2,3, 
\la{fKappa}
\ee
with $\kappa$ being the deformation parameter of the dimension of length\footnote{ {Throughout this article we follow the conventions of~\cite{Basilio:2024bir}, which differ from those of~\cite{Lukierski:1992dt}, where the deformation parameter is denoted by $1/\kappa$ and the commutative limit is achieved at $\kappa\to\infty$. }}. 
The corresponding structure constants are
\be
f^{\mu\nu}_{\xi} = \kappa\,(\delta_0^{\mu}\delta^{\nu}_{\xi} - \delta_0^{\nu}\delta^{\mu}_{\xi}). 
\ee

The local components $\gamma^\mu_\nu (p)$ and $\rho^\mu_\nu (p)$ of the left-invariant vector fields and right-invariant one-forms are given by the matrices\footnote{According to our notations, for any matrix, the upper index enumerates the rows, while the lower index enumerates the columns.}
\be
\gamma(p)  =  \left(
\begin{array}{cccc}
1 &-\kappa p_1 &-\kappa p_2 & -\kappa p_3 \\
0 &1 &0 &0 \\
0 &0 &1 &0 \\
0 &0 &0 &1 
\end{array}
\right), 
\qquad 
\rho(p)  =  \left(
\begin{array}{cccc}
1 &0 &0 & 0 \\
0 & e^{\kappa p_0} &0 &0 \\
0 &0 &e^{\kappa p_0} &0 \\
0 &0 &0 &e^{\kappa p_0} 
\end{array}
\right). \la{newRG}
\ee

 We shall work with the gauge-covariant field equations proposed in~\cite{Kupriyanov:2023gjj}:
\be
\mathcal{E}^\mu = 0,
\qquad
\mathcal{E}^\mu :=   
 \mathcal{D}_\xi \mathcal{F}^{\xi\mu} + \frac{1}{2}\, \mathcal{F}_{\zeta \phi}\,f_\nu^{\zeta \phi}\, \mathcal{F}^{\mu\nu} -\mathcal{F}_{\zeta \phi} \, f_\nu^{\mu \phi}\,\mathcal{F}^{\zeta \nu}  -\frac{1}{12}\,\big( f_{\nu}^{\nu \mu}\, \mathcal{F}_{\phi \zeta}\,\mathcal{F}^{\phi \zeta} 
+ 4\,f_\nu^{\nu \phi}\,\mathcal{F}_{\phi \zeta}\,\mathcal{F}^{\zeta \mu}\big), \la{kGeom}
\ee
where
\be
\mathcal{F}^{\mu\nu} = \eta^{\mu \xi}\,\eta^{\nu \zeta}\, \mathcal{F}_{\xi \zeta},
\ee
and the metric tensor $\eta$, used to raise indices, is the flat Minkowski metric:
\be
\eta = \mathrm{diag}\,(\,+1,-1,\,-1,\,-1). \la{eta}
\ee

The first term in $\mathcal{E}^\mu$ recovers the left-hand side of the Maxwell equations in the commutative limit, while the remaining terms, which vanish at  $\kappa \to 0$, are needed to maintain the gauge identity
\be
\mathcal{D}_\mu\mathcal{E}^\mu = 0.
\ee

\section{Localized signals}
\la{SOLU}
The classical gauge potentials,
\be
A^{\mathrm{cl} }_{\pm}(x) = \big(v(s_{\pm}), w(s_{\pm}), 0, 0\big),
\ee
with
\be
s_{\pm} := x^3 \pm x^0,
\ee
satisfy the usual Maxwell equations for any sufficiently smooth form factors $v(s)$ and $w(s)$.  When the supports of $v(s)$  and  $w(s)$ are compact, these solutions describe superpositions of localized wave packets, which are polarized along the $x^1$ and $x^2$ axes, respectively. Both wave packets propagate along the $x^3$-axis (or in the opposite direction) at the speed of light, which is equal to one in natural units.

 Remarkably, these field configurations exhibit noncommutative analogs. Indeed, one can verify by direct substitution that the following one-parameter family of gauge potentials:
\be
A_{\pm}(x;\alpha) = \bigg(\alpha\,v\big(s_{\pm}^{\alpha}\big),\alpha\,w\big(s_{\pm}^{\alpha}\big), \pm\frac{\alpha}{\kappa} \,\Big(1-\sqrt{1-\kappa^2\big[\big(v^2 (s^{\alpha}_{\pm})+w^2 (s^{\alpha}_{\pm})\big]}\Big), -\frac{\ln{(\alpha)}}{\kappa}\bigg), 
\label{sols}
\ee
with
\be
s_{\pm}^{\alpha} :=\alpha x^3 \pm x^0, \qquad \forall\alpha>0,
\ee
satisfy the noncommutative field equations~\eqref{kGeom}. The fields  $A_{\pm}$ are real-valued iff the form factors obey the inequality
\be
v^2 + w^2 \leq \frac{1}{\kappa^2}.
\ee
In what follows, we assume that this condition is fulfilled. 

The corresponding components of the noncommutative field strength $\mathcal{F}_{\mu\nu}$  for  $\mu > \nu$ are given by  
\bea 
\mathcal{F}^{\pm}_{10}(x;\alpha)&=& \mp v^{'} ( s_{\pm}^{\alpha} ), \nonumber\\
\mathcal{F}^{\pm}_{20}(x;\alpha)&=& \mp w^{'} ( s_{\pm}^{\alpha} ),  \nonumber\\
\mathcal{F}^{\pm}_{30}(x;\alpha)&=& -{\frac {\kappa\, \big( w ( s_{\pm}^{\alpha} ) w^{'} ( s_{\pm}^{\alpha} )
 +v ( s_{\pm}^{\alpha} ) v^{'} ( s_{\pm}^{\alpha} )  \big) }{
 \sqrt{1-{\kappa}^{2} \big(v^2 ( s_{\pm}^{\alpha} )  +  w^2 ( s_{\pm}^{\alpha} ) \big) }
 }},
 \nonumber\\
\mathcal{F}^{\pm}_{21}(x;\alpha)&=&
\pm\kappa\, \big(   w^{'} ( s_{\pm}^{\alpha} )   v ( s_{\pm}^{\alpha} ) -w ( s_{\pm}^{\alpha} ) v^{'} ( s_{\pm}^{\alpha} ) \big),\nonumber\\
\mathcal{F}^{\pm}_{31}(x;\alpha)&=&{\frac { v^{'} ( s_{\pm}^{\alpha} )+ {\kappa}^{2} w ( s_{\pm}^{\alpha} )\big( w^{'} ( s_{\pm}^{\alpha} )    v ( s_{\pm}^{\alpha} ) 
-  w ( s_{\pm}^{\alpha} )   v^{'} ( s_{\pm}^{\alpha} )   \big)
 }{
 \sqrt{1-{\kappa}^{2} \big(v^2 ( s_{\pm}^{\alpha} )  +  w^2 ( s_{\pm}^{\alpha} ) \big) }
}},
\nonumber\\
\mathcal{F}^{\pm}_{32}(x;\alpha) &=&{\frac {w^{'} ( s_{\pm}^{\alpha} )
+ {\kappa}^{2}v ( s_{\pm}^{\alpha} ) \big( v^{'} ( s_{\pm}^{\alpha} )    w ( s_{\pm}^{\alpha} ) 
-  v ( s_{\pm}^{\alpha} )   w^{'} ( s_{\pm}^{\alpha} )   \big)
 }{
 \sqrt{1-{\kappa}^{2} \big(v^2 ( s_{\pm}^{\alpha} )  +  w^2 ( s_{\pm}^{\alpha} ) \big) } 
 }} . \la{Fs}
\eea

For any fixed value of the parameter  $\alpha$, the group velocity, that is, the speed of light, is equal to
\begin{equation}
v_{\mathrm{light}}=\alpha^{-1}.
\label{lightspeed}
\end{equation}
In other words, it can take any value! Interestingly, though the expressions~\eqref{Fs} exhibit finite commutative limits, the corresponding gauge potentials~\eqref{sols} are singular at $\kappa \to 0$ , unless $\alpha = 1$. Therefore, the ambiguity of the speed of light is a non-perturbative phenomenon which arises from the noncommutativity of the gauge algebra.  

In the next section, we take a closer look at this arbitrariness by establishing an important fact: the potentials  $A_{\pm}(x, \alpha)$ for different values of $\alpha$ are related by (finite) Poisson gauge transformations. Since gauge transformations always map solutions of the field equations onto other solutions, the $\alpha$-arbitrariness is mathematically expected. However, its physical significance is more subtle and will be discussed in Sec.~\ref{WGO}.
 
\section{Finite gauge transformations}
\la{GT}
In Sec.~\ref{formalism}, we introduced the infinitesimal Lie-Poisson gauge transformations~\eqref{igt}. Below, we describe the general construction of their finite counterparts, proposed in~\cite{Kupriyanov:2023qot}, and apply it to our specific case. For practical purposes, we work exclusively in local coordinates.

The prescription of~\cite{Kupriyanov:2023qot} is based on three key technical ingredients: the group operation law in local coordinates, the coadjoint action of the group $G$ on the space-time, and the pure gauge potentials. We introduce each of these elements separately and then combine them.

\subsubsection*{a. The group operation law in local coordinates}
Let $\oplus$ be the group operation law on $G$, expressed in local coordinates;  {that is, a binary operation on the coordinate chart covering $G$,
\be
(p,q)\mapsto p\oplus q,\qquad  p,q  \in \mathbb{R}^d,
\ee
such that:}
\be
g(p)\cdot g(q)=g(p\oplus q).
\ee
For the $\kappa$-Minkowski noncommutativity~\eqref{fKappa}, the group $G$ is the four-dimensional space $\mathbb{R}^{4}$, equipped with the following group operation:
\bea
(p \oplus q)_a &=&   e^{-\kappa q_0} p_a + q_a, \qquad a=1,2,3, \nonumber \\
(p \oplus q)_0 &=& p_0 +  q_0. \la{composition}
\eea
One can easily check that all group axioms hold, and the relations
\be
\gamma^\mu_\nu(p) = \frac{\partial}{\partial q_\mu} \big(p\oplus q\big)_\nu\bigg|_{q=0}, \qquad \bar\rho^\mu_\nu(p) = \frac{\partial}{\partial q_\mu} \big(q\oplus p\big)_\nu\bigg|_{q=0}, \la{rele}
\ee  
correctly reproduce the left-invariant vector fields and right-invariant one-forms~\eqref{newRG} on $G$.  In~\eqref{rele},  $\bar\rho$  denotes the inverse matrix of $\rho$.

\subsubsection*{b. Coadjoint action of $G$ on the space-time}
By identifying the space-time  $\mathcal{M} \simeq \mathbb{R}^d$ with the Lie coalgebra $\mathfrak{g}^*\simeq \mathbb{R}^d $ of the group $G$, it is natural to introduce the following diffeomorphisms of $\mathcal{M}$:
\be
x\longrightarrow x_{(p)} :=\mathrm{Ad}^*_{g(p)}(x).
\la{xcoadj}
\ee
In the $\kappa$-Minkowski case, we get:
\bea
x^a_{(p)}  &=& e^{-\kappa p_0} x^a,\qquad a=1,2,3 \nonumber\\
x^0_{(p)} &=& x^0 + \kappa\,(p_1\, x^1 + p_2\, x^2 + p_3\, x^3). \la{coadjour}
\eea 

\subsection*{c. Pure gauge potentials}
By definition, the pure gauges are the gauge potentials $\Lambda(x)$, which satisfy the equation
\be
\bar{\gamma}^{\mu}_\sigma(\Lambda) \partial_\nu \Lambda_\mu - \bar{\gamma}^{\mu}_\nu(\Lambda) \partial_\sigma \Lambda_\mu
+ \frac{1}{2}\bar{\gamma}^\mu_\xi(\Lambda) f^{\xi \chi}_\omega\, x^\omega\, \bar{\gamma}^\zeta_\chi(\Lambda)\, \partial_\nu \Lambda_\mu \,\partial_\sigma \Lambda_\zeta =0, 
\la{LagrSec}
\ee
where $\bar\gamma$ stands for the inverse matrix of $\gamma$.
 As in the commutative case, any constant field satisfies this relation. Below, we shall work with the following one-parameter family of pure gauges:
\be
\Lambda(x;\beta) = \bigg(\,\,\, 0,\,\,\, 0,\,\,\, 0,\,\,\,-\frac{\ln{(\beta)}}{\kappa} \,\,\,\bigg), \qquad \forall \beta > 0 .\la{ourpug}
\ee

\subsection*{d. All together}
For a given pure gauge potential $\Lambda$, the finite Lie-Poisson gauge transformation is defined as follows:
\be
A(x)\longrightarrow \tilde{A}(x):=A\big(x_{(p)}\big) \big|_{p=\Lambda(x)}\oplus \Lambda(x).
\la{fgt}
\ee
The corresponding infinitesimal gauge transformations are exactly~\eqref{igt}. Indeed, for small $\Lambda$ the matrix $\gamma$ tends to the identity matrix, thus in this approximation the pure gauge condition~\eqref{LagrSec} reduces to
\be
 \partial_\nu \Lambda_\sigma - \partial_\sigma \Lambda_\nu = 0,
\ee
where we neglected  {quadratic terms} in $\Lambda$.
Therefore, infinitesimal pure gauge potentials take the ‘usual’ form $\Lambda_{\mu} = \partial_{\mu} f$ for  {a} smooth function $f$, which plays the role of the infinitesimal gauge parameter. 
By substituting this expression into~\eqref{fgt} and calculating the first variation over $f$, one arrives at~\eqref{igt}.

The finite gauge transformations~\eqref{fgt} naturally arise from the groupoid formalism (see~\cite{Kupriyanov:2023qot} for details). Generic gauge fields correspond to bisections of a proper symplectic groupoid, while pure gauge potentials are related to Lagrangian bisections. All bisections, equipped with a suitable multiplication rule, form a group. The finite gauge transformation~\eqref{fgt} of a given gauge potential $A$ is simply the right multiplication of the corresponding bisection by a Lagrangian bisection representing the pure gauge $\Lambda$.

An important novelty of the transformations~\eqref{fgt}, compared to the commutative case, is the presence of the diffeomorphisms~\eqref{xcoadj}. As we shall soon see, these diffeomorphisms are closely related to the gauge-dependence of the speed of light. On the other hand, as we shall explain in the next section, the Poisson gauge transformations act as diffeomorphisms on the configuration space of the charged particle as well. These two actions compensate each other in such a way that the results of measurements remain gauge-invariant. 

For our pure gauge~\eqref{ourpug}, the diffeomorphisms~\eqref{xcoadj} at $p= \Lambda$, which enter the right-hand side of Eq.~\eqref{fgt}, reduce to simple rescalings of the spatial coordinates:
\be
x_{(p)}^0\big|_{p=\Lambda} = x^0, \qquad  x_{(p)}^j\big|_{p=\Lambda} = \beta\, x^j. \la{ourcoe}
\ee
By substituting our data~\eqref{sols} and~\eqref{ourpug} into the general expression~\eqref{fgt} for a finite gauge transformation and by using the group operation~\eqref{composition} together with the coadjoint action~\eqref{ourcoe}, we arrive at the following finite gauge transformations:
\be
A_{\pm}(x;\alpha)\longrightarrow\tilde A_{\pm}(x;\alpha) =  A_{\pm}(x;\alpha\cdot\beta) , \la{transformedpotrntial}
\ee 
which map the fields $A_{\pm}(x;\alpha)$ with different values of $\alpha$ onto each other.

Remarkably, the pure gauge potentials~\eqref{ourpug} are singular at $\kappa\to 0$. The solutions $A_{\pm}(x;\alpha)$ can be obtained through a singular gauge transformation from the fields $A_{\pm}(x;1)$, which have a well-defined commutative limit. Therefore, it is not surprising that $A_{\pm}(x;\alpha)$ do not exhibit a smooth commutative limit. 

The relation~\eqref{lightspeed} implies the following transformation rule for the speed of light under our gauge transformations:
\be
v_{\mathrm{light}} \longrightarrow \frac{1}{\beta}\, v_{\mathrm{light}},  \la{slatransform}
\ee
which may give the impression that different gauges are physically inequivalent. In the next section, we explain why this naive conclusion is wrong. Moreover, we will show that the gauge-dependent speed of light is unavoidable for the physical consistency of the formalism.

\section{On the physical equivalence of various gauges }
\la{WGO}
To measure the speed of light, one must consider the interaction between the electromagnetic field and charged matter, which constitutes the experimental devices, so we take a closer look at the motion of charged particles.

Let $\tau$ be a parameter on the world-line of a test particle. A peculiar feature of the Lie-Poisson gauge formalism is the gauge dependence of the particle’s trajectory. Upon an infinitesimal Poisson gauge transformation of the gauge background~\eqref{igt}, the particle’s position  $x^{\mu}(\tau)$ transforms in a non-trivial way:
\be
x^{\mu}(\tau)\longrightarrow \tilde{x}^{\mu}(\tau) := x^{\mu}(\tau) + \delta_{f}x^{\mu}(\tau),
\ee
with
\be
\delta_{f}x^{\mu}(\tau) = \left\{f(y),y^{\mu}\right\}_{y=x(\tau)}. \la{igtp}
\ee
These gauge transformations map the solutions $x(\tau)$ of the equations of motion for charged particles in the original gauge background $A$ to solutions  $\tilde{x}(\tau)$ of the equations of motion in the transformed gauge background $\tilde{A}$. Thus, Poisson gauge transformations map one dynamics of all fields and particles in the world onto another. This property conceptually distinguishes the noncommutative formalism from the usual $U(1)$  gauge theory, where different gauges provide alternative descriptions of the same dynamics. The particle’s equations of motion and all relevant details are presented in~\cite{Basilio:2024bir}.

For a finite Poisson gauge transformation~\eqref{fgt} of $A_{\mu}(x)$, the corresponding finite transformation of the particle’s world-line
\be
x^{\mu}(\tau)\longrightarrow \tilde{x}^{\mu}(\tau).
\ee 
can be obtained by solving the algebraic equation
\be
x=\mathrm{Ad}^*_{g\left(\Lambda(\tilde{x})\right)}(\tilde{x}),
\ee
with respect to $\tilde{x}$; see~\cite{KupriyanovUnp} for details. For the pure gauge potential~\eqref{ourpug}, the solutions are given by
\be
\tilde{x}^0 =  x^0,\qquad \tilde{x}^j = \beta^{-1}\, x^j. \la{xtrans}
\ee

The gauge transformations, which rescale the speed of light, also rescale all distances between matter particles in the same way. Therefore, the time required for a signal to travel from one point in space to another one remains invariant under these gauge transformations.

To measure a physical quantity means to compare it with the corresponding reference value, called a unit of measurement. In the International System of Units (SI), the unit of measurement of length, that is, one meter is defined as the distance traveled by an electromagnetic signal in vacuum during a time interval of $1/299792458$ seconds:
\be
1\, \mathrm{meter}  = v_{\mathrm{light}}\,\frac{1}{299792458}\, 1 \,\mathrm{second}. \la{meter}
\ee
When we perform our gauge transformations, by rescaling the speed of light~\eqref{slatransform}, we rescale this reference distance:
 \be
 1\, \mathrm{meter} \longrightarrow \frac{1}{\beta}\, 1\, \mathrm{meter}.  \la{onemtra}
 \ee
However, according to~\eqref{xtrans}, we also rescale the spatial distance $L$ that we intend to measure:
\be
L \longrightarrow \frac{1}{\beta}\, L.
\ee
Therefore, the measured length in meters, that is, the dimensionless ratio $L/\mathrm{meter}$, remains exactly the same. In this sense, all gauges with different values of $v_{\mathrm{light}}$ are physically equivalent. 

Of course, the same conclusion applies to the speed of light.  
Indeed, according to~\eqref{slatransform} and~\eqref{onemtra} the measured value of the speed of light, that is, the dimensionless ratio 
$$
v_{\mathrm{light}}/( \mathrm{meter}\cdot\mathrm{second}^{-1})
$$
remains invariant under our gauge transformations and equals $299792458$. We emphasize that the time-intervals remain unchanged under the gauge transformations~\eqref{xtrans}. At the end of the day, the gauge ambiguity of the parameter $v_{\mathrm{light}}$, established in the previous section, is absorbed into the redefinition of 1 meter (cf. Eq.~\eqref{meter}) and cannot be measured.

The gauge-dependence~\eqref{slatransform} of the parameter $v_{\mathrm{light}}$ is actually a desired property. Indeed, as we have seen, our gauge transformations rescale all the spatial distances between the particles. The transformed dynamics would be equivalent to the original one iff the speed of light scales exactly as the distances. Otherwise we would run into the following paradox. Consider two experimentalists Alice and Bob, and a signal, which Alice sends to Bob. Let $L$ be the distance between them. Bob will detect the signal after the time-interval $\Delta t = L/v_{\mathrm{light}}$. By setting $\beta$, say, equal to $1/100$ we map the original dynamics onto the new one, where the distance $L$ between Alice and Bob is $100$ times larger. If the speed of light was gauge invariant, after the same time interval, the signal would pass just one percent of the whole distance, so Bob’s detector would not detect anything!  Fortunately, the gauge-dependence~\eqref{slatransform} of the speed of light allows us to avoid this pathology. By increasing the speed of light $100$ times, the transformed dynamics guarantees that Bob will detect the signal after the time interval $\Delta t$.

In conclusion, we comment on the gauge-invariant combinations of dynamical variables in this formalism. Consider the components of the deformed field strength evaluated at the test particle’s position, $\mathcal{F}_{\mu\nu}\big(x(\tau)\big)$. As was noticed in~\cite{Basilio:2024bir}, the transformation laws~\eqref{igtF} and~\eqref{igtp} for $\mathcal{F}_{\mu\nu}(x)$ and $x(\tau)$ imply that,
\be
\delta_{f}\,\mathcal{F}_{\mu\nu}\big(x(\tau)\big) = 0,
\ee
so we conclude that $\mathcal{F}(x(\tau))$ is gauge-invariant under noncommutative gauge transformations:
\be
\tilde{\mathcal{F}}_{\mu\nu}\big( \tilde{x}(\tau)\big) = \mathcal{F}_{\mu\nu}\big( {x}(\tau)\big). \la{Fpgainv}
\ee
Of course, the same logic is applicable not only to $\mathcal{F}$ but to its gauge-covariant derivatives of any order as well,
\be
\tilde{\mathcal{D}}_{\xi_1}\cdots \tilde{\mathcal{D}}_{\xi_n}\,\tilde{\mathcal{F}}_{\mu\nu}\big( \tilde{x}(\tau)\big) =
{\mathcal{D}}_{\xi_1}\cdots {\mathcal{D}}_{\xi_n}\,{\mathcal{F}}_{\mu\nu}\big( {x}(\tau)\big).
\ee
In this equality  $\tilde{\mathcal{D}}$ denotes the gauge-covariant derivative~\eqref{gcodder}, constructed with the gauge transformed field $\tilde{A}$.

\section{Summary and concluding discussions}
\la{conclu}
We studied signal propagation in the Lie-Poisson electrodynamics with the $\kappa$-Minkowski noncommutativity. Despite the complexity and nonlinearity of the deformed Maxwell equations~\eqref{kGeom}, we succeeded to construct a one-parameter family of the exact solutions~\eqref{sols}, which describe the wave packets with arbitrary polarizations and profiles, which travel along a given direction.

The solutions characterized by different values of the parameter  $\alpha$ correspond to signals traveling with the group velocity $1/\alpha$. All these solutions are related to each other by finite gauge transformations, what implies a gauge dependence of the speed of light.

On the other hand, the trajectories of particles are gauge-dependent as well. In conventional commutative $U(1)$ gauge theory, gauge transformations provide different descriptions of the same dynamics. In contrast, in the Lie-Poisson gauge formalism, gauge transformations map one dynamics of fields and particles onto another. In particular, according to Eq.~\eqref{xtrans}, the gauge transformations that rescale the speed of light also rescale all distances between particles.

The gauge dependence of both the speed of light and the distances between charged particles is self-consistent in the following sense. Under the gauge transformations~\eqref{transformedpotrntial} and~\eqref{xtrans}, which affect both fields and particle positions, the time taken by a signal to travel from one particle to another remains independent of the parameter $\beta$ . By defining the unit of length in the standard way~\eqref{meter}, we see that observers associated with gauge-related realizations of the dynamics will measure the same length and the same speed of light. The gauge independence of measurement results implies that gauges characterized by different values of $\alpha$ are physically equivalent.

In conclusion, we comment on the model independence. The discussion in Sec.~\ref{GT} and Sec.~\ref{WGO} concerns gauge transformations alone and does not rely on any particular choice of a dynamical model. It is applicable to any dynamical model that exhibits traveling wave packet solutions. Of course, all these considerations make sense iff these solutions of the noncommutative field equations do exist. The analysis in Sec.~\ref{SOLU} demonstrates that traveling wave packets can indeed be realized within the Lie-Poisson gauge formalism and $\kappa$-Minkowski noncommutativity.

The formalism, designed so far, contains the electromagnetic field and the charged particles only, and the theory has not yet been quantized. A suitable quantization of the theory, along with the addition of other interactions, e.g. the gravitational one, will not alter our conclusions provided the Lie-Poisson gauge symmetry is respected.

\section*{Acknowledgements}
The author is grateful to Vlad Kupriyanov, Fedele Lizzi, Alexey Sharapov, and Patrizia Vitale for valuable discussions on this and related papers.


\begin{thebibliography}{99}

\bibitem{Seiberg:1999vs}
N.~Seiberg and E.~Witten,
``String theory and noncommutative geometry,''
JHEP \textbf{09} (1999), 032
doi:10.1088/1126-6708/1999/09/032


\bibitem{Freidel:2005me}
L.~Freidel and E.~R.~Livine,
``3D Quantum Gravity and Effective Noncommutative Quantum Field Theory,''
Phys. Rev. Lett. \textbf{96} (2006), 221301
doi:10.1103/PhysRevLett.96.221301

\bibitem{Doplicher:1994tu}
S.~Doplicher, K.~Fredenhagen and J.~E.~Roberts,
``The Quantum structure of space-time at the Planck scale and quantum fields,''
Commun. Math. Phys. \textbf{172} (1995), 187-220

\bibitem{Szabo} R.~J.~Szabo,
``Quantum field theory on noncommutative spaces,''
Phys. Rept. \textbf{378} (2003), 207-299
doi:10.1016/S0370-1573(03)00059-0

\bibitem{Hersent:2022gry}
K.~Hersent, P.~Mathieu and J.~C.~Wallet,
``Gauge theories on quantum spaces,''
Phys. Rept. \textbf{1014} (2023), 1-83
doi:10.1016/j.physrep.2023.03.002



\bibitem{Kupriyanov:2019ezf}  
V.~G.~Kupriyanov,
``$L_\infty$-Bootstrap Approach to Non-Commutative Gauge Theories,''
Fortsch. Phys. \textbf{67} (2019) no.8-9, 1910010
doi:10.1002/prop.201910010

\bibitem{Kupriyanov:2019cug} 
V.~G.~Kupriyanov,
``Non-commutative deformation of Chern\textendash{}Simons theory,''
Eur. Phys. J. C \textbf{80} (2020) no.1, 42
doi:10.1140/epjc/s10052-019-7573-y

\bibitem{Kupriyanov:2020sgx}  
V.~G.~Kupriyanov and P.~Vitale,
``A novel approach to non-commutative gauge theory,''
JHEP \textbf{08} (2020), 041
doi:10.1007/JHEP08(2020)041


\bibitem{Kupriyanov:2020axe}  
V.~G.~Kupriyanov, M.~Kurkov and P.~Vitale,
``$\kappa$-Minkowski-deformation of U(1) gauge theory,''
JHEP \textbf{01} (2021), 102
doi:10.1007/JHEP01(2021)102


\bibitem{Kupriyanov:2021aet} 
V.~G.~Kupriyanov,
``Poisson gauge theory,''
JHEP \textbf{09} (2021), 016
doi:10.1007/JHEP09(2021)016


\bibitem{Kupriyanov:2021cws}
V.~G.~Kupriyanov and R.~J.~Szabo,
``Symplectic embeddings, homotopy algebras and almost Poisson gauge symmetry,''
J. Phys. A \textbf{55} (2022) no.3, 035201
doi:10.1088/1751-8121/ac411c



\bibitem{Abla:2022wfz}
O.~Abla, V.~G.~Kupriyanov and M.~A.~Kurkov,
``On the L$_{\infty}$ structure of Poisson gauge theory,''
J. Phys. A \textbf{55} (2022) no.38, 384006
doi:10.1088/1751-8121/ac87df



\bibitem{Kupriyanov:2022ohu}
V.~G.~Kupriyanov, M.~A.~Kurkov and P.~Vitale,
``Poisson gauge models and Seiberg-Witten map,''
JHEP \textbf{11} (2022), 062
doi:10.1007/JHEP11(2022)062


\bibitem{Abla:2023odq}
O.~Abla and M.~J.~Neves,
``Effects of wave propagation in canonical Poisson gauge theory under an external magnetic field,''
EPL \textbf{144} (2023) no.2, 24001
doi:10.1209/0295-5075/ad0574

\bibitem{Kupriyanov:2023gjj}
V.~G.~Kupriyanov, M.~A.~Kurkov and P.~Vitale,
``Lie-Poisson gauge theories and \ensuremath{\kappa}-Minkowski electrodynamics,''
JHEP \textbf{11} (2023), 200
doi:10.1007/JHEP11(2023)200



\bibitem{Kupriyanov:2023qot}
V.~G.~Kupriyanov, A.~A.~Sharapov and R.~J.~Szabo,
``Symplectic groupoids and Poisson electrodynamics,''
JHEP \textbf{03} (2024), 039
doi:10.1007/JHEP03(2024)039


\bibitem{DiCosmo:2023wth}
F.~Di Cosmo, A.~Ibort, G.~Marmo and P.~Vitale,
``Symplectic realizations and Lie groupoids in Poisson Electrodynamics,''
[arXiv:2312.16308 [hep-th]].


\bibitem{Bascone:2024mxs}
F.~Bascone and M.~Kurkov,
``Hamiltonian analysis in Lie\textendash{}Poisson gauge theory,''
Int. J. Geom. Meth. Mod. Phys. \textbf{21} (2024) no.06, 2450108
doi:10.1142/S0219887824501081


\bibitem{Sharapov:2024bbu}  
A.~A.~Sharapov,
``Poisson electrodynamics with charged matter fields,''
J. Phys. A \textbf{57} (2024) no.31, 315401
doi:10.1088/1751-8121/ad62c7

\bibitem{Kupriyanov:2024dny}
V.~Kupriyanov, M.~Kurkov and A.~Sharapov,
``Classical mechanics in noncommutative spaces: confinement and more,''
Eur. Phys. J. C \textbf{84} (2024) no.10, 1068
doi:10.1140/epjc/s10052-024-13372-7

\bibitem{Basilio:2024bir}
B.~S.~Basilio, V.~G.~Kupriyanov and M.~A.~Kurkov,
``Charged particle in Lie\textendash{}Poisson electrodynamics,''
Eur. Phys. J. C \textbf{85} (2025) no.2, 175
doi:10.1140/epjc/s10052-025-13897-5


\bibitem{Abla:2024wtr}
O.~Abla and M.~J.~Neves,
``Poisson electrodynamics on {\ensuremath{\kappa}}-Minkowski space-time,''
Phys. Lett. B \textbf{864} (2025), 139385
doi:10.1016/j.physletb.2025.139385






\bibitem{Lizzi:2020tci}
F.~Lizzi, M.~Manfredonia and F.~Mercati,
``The momentum spaces of $\kappa$-Minkowski noncommutative spacetime,''
Nucl. Phys. B \textbf{958} (2020), 115117
doi:10.1016/j.nuclphysb.2020.115117

\bibitem{Mathieu:2020ywc}
P.~Mathieu and J.~C.~Wallet,
``Single Extra Dimension from $\kappa$-Poincar\'e and Gauge Invariance,''
JHEP \textbf{03} (2021), 209
doi:10.1007/JHEP03(2021)209


\bibitem{Lizzi:2021rlb}
F.~Lizzi and F.~Mercati,
``$\kappa$-Poincar\'e-comodules, Braided Tensor Products and Noncommutative Quantum Field Theory,''
Phys. Rev. D \textbf{103} (2021), 126009
doi:10.1103/PhysRevD.103.126009




\bibitem{Mathieu:2021mxl}
P.~Mathieu and J.~C.~Wallet,
``Twisted BRST symmetry in gauge theories on the $\kappa$-Minkowski spacetime,''
Phys. Rev. D \textbf{103} (2021) no.8, 086018
doi:10.1103/PhysRevD.103.086018

\bibitem{Meljanac:2022qed} 
S.~Meljanac, Z.~\v{S}koda and S.~Kresic-Juric,
``Symmetric ordering and Weyl realizations for quantum Minkowski spaces,''
J. Math. Phys. \textbf{63} (2022) no.12, 123508
doi:10.1063/5.0094443

\bibitem{Carotenuto:2020vnv}
A.~Carotenuto, F.~Lizzi, M.~Manfredonia and F.~Mercati,
``The Weyl\textendash{}Mellin quantization map,''
Int. J. Geom. Meth. Mod. Phys. \textbf{19} (2022) no.03, 2250031
doi:10.1142/S0219887822500311


\bibitem{Fabiano:2023xke}
G.~Fabiano and F.~Mercati,
``Multiparticle states in braided lightlike {\ensuremath{\kappa}}-Minkowski noncommutative QFT,''
Phys. Rev. D \textbf{109} (2024) no.4, 046011
doi:10.1103/PhysRevD.109.046011

\bibitem{Lukierski:1992dt}
J.~Lukierski, A.~Nowicki and H.~Ruegg,
``New quantum Poincare algebra and k deformed field theory,''
Phys. Lett. B \textbf{293} (1992), 344-352
doi:10.1016/0370-2693(92)90894-A

\bibitem{KupriyanovUnp} 
F.~Di Cosmo, V.~G.~Kupriyanov and P.~Vitale,
``The electromagnetic field in Poisson gauge theory: the groupoidal approach,''
[arXiv:2510.04858 [hep-th]].




\end{thebibliography}
\end{document}